\documentclass[a4paper]{jpconf}
\usepackage{graphicx}
\begin{document}
\title{Parametric instabilities in the LCGT arm cavity}

\author{K Yamamoto$^1$,
T Uchiyama$^1$, S Miyoki$^1$, M Ohashi$^1$, K Kuroda$^1$,\\ 
K Numata$^2$}

\address{$^1$ Institute for Cosmic Ray Research, The University of Tokyo, 
Kashiwa, Chiba 277-8582, Japan}
\address{$^2$ NASA Goddard Space Flight Center, CRESST, Code663, Greenbelt, 
MD 20771, U.S.A.}

\ead{yamak@icrr.u-tokyo.ac.jp}

\begin{abstract}
We evaluated the parametric instabilities of LCGT 
(Japanese interferometric gravitational wave 
detector project) arm cavity.  
The number of unstable modes of 
LCGT is 10-times smaller than that of Advanced LIGO (U.S.A.). 
Since the strength of the instabilities of LCGT 
depends on the mirror curvature   
more weakly than that of Advanced LIGO, 
the requirement of the mirror curvature accuracy is easier to be achieved. 
The difference in the parametric instabilities between LCGT 
and Advanced LIGO
is because of the thermal noise reduction methods 
(LCGT, cooling sapphire mirrors; 
Advanced LIGO, fused silica mirrors with larger laser beams), 
which are the main strategies of the projects.   
Elastic Q reduction by the barrel surface 
(0.2 mm thickness Ta$_2$O$_5$) coating 
is effective to suppress instabilities in the LCGT arm cavity.
Therefore, the cryogenic interferometer is a smart solution 
for the parametric instabilities in addition 
to thermal noise and thermal lensing. 
\end{abstract}

\section{Introduction}

Observations using several interferometric gravitational wave detectors 
(LIGO \cite{LIGO}, 
Virgo \cite{VIRGO}, GEO \cite{GEO}, TAMA \cite{TAMA}) on the ground 
are presently 
in progress. 
In order to construct detectors with
better sensitivity, 
future (second generation) projects 
have been proposed: 
Advanced LIGO (U.S.A.) \cite{advancedLIGO} and  
LCGT (Japan) \cite{LCGT}. These projects have km-scale Fabry-Perot cavities. 
The optical transversal mode spacing in these long 
cavities and the intervals of the elastic modes of 
the mirrors 
are on the order of 10 kHz. When the optical transversal mode spacing 
are comparable with the intervals of 
the elastic modes, parametric instabilities become a problem 
in the stable operation of 
the interferometer \cite{Braginsky1}. 
Small thermally driven elastic vibration modulates 
the light and excites the 
transverse modes of the cavity. These excited optical modes apply 
modulated radiation pressure on the mirrors. This 
makes the amplitude of the elastic modes larger. 
At last, the elastic modes and optical modes, 
except for TEM00, oscillate largely. 

The formula of the parametric instability (without power recycling, 
resonant sideband extraction, 
and anti-Stokes modes) is derived in Ref. \cite{Braginsky1}. 
If a parameter, $R$, of an elastic 
mode is larger than unity, that mode is unstable. 
The formula of $R$ is  
\begin{equation}
R = \sum_{\rm optical\ mode} \frac{4PQ_{\rm m}Q_{\rm o}}{McL{\omega_{\rm m}}^2}
\frac{\Lambda_{\rm o}}{1+\Delta \omega^2/{\delta_{\rm o}}^2},
\label{R}
\end{equation}  
where $P,Q_{\rm m},Q_{\rm o},M,c,L,\omega_{\rm m},\Delta\omega$, 
and $\delta_{\rm o}$ 
are the optical power in the cavity, 
the Q-values of the elastic and optical modes, the mass of the mirror, 
the speed of light, the 
cavity length, the angular frequency of the elastic mode, 
the angular frequency differences between the 
elastic and optical modes, and the half-width angular frequency 
of the optical mode, 
respectively. 
The value $\Lambda_{\rm o}$
represents the spatial overlap between the optical and elastic modes. 
If the shapes of the optical 
and elastic modes are similar, $\Lambda_{\rm o}$ is on the order of unity. 
If the shapes are not similar, $\Lambda_{\rm o}$ is 
almost zero. When the shapes and frequencies of the optical 
and elastic modes are similar 
($\Lambda_{\rm o} \sim 1, \Delta \omega \sim 0$), $R$ will become 
several thousand in future projects \cite{PR}. 
These parametric instabilities are a serious problem in Advanced 
LIGO \cite{Ju1,Ju2}. The instabilities of the LCGT interferometer 
have never been considered. 
We evaluated the instabilities of the LCGT arm cavity. 
In Sec. 2, we introduce the calculation results 
for the parametric instabilities
in LCGT and compare 
them with those in Advanced LIGO.
In order to calculate $\omega_{\rm m}$ and 
$\Lambda_{\rm o}$ for the instability evaluation, 
we used ANSYS, which is a software application 
for a finite-element method \cite{ANSYS}. In Sec. 3, the difference in the 
parametric instabilities between Advanced LIGO and LCGT is discussed. 
In Sec. 4, how to 
suppress the instabilities of the LCGT arm cavity is considered. 
Sec. 5 and Sec. 6 are
devoted to future work and a summary, respectively.   

\section{Results of calculation}

\subsection{Specification}

Table \ref{specification} gives the specifications of 
Advanced LIGO in Refs. \cite{Ju1,Ju2,Zhao} (after these references, 
the specifications of Advanced LIGO were changed slightly)
and LCGT \cite{Kuroda} (the exact values of the LCGT mirror 
curvature are not fixed. 
The curvature given in Table \ref{specification} is only a candidate). 
The important differences between 
Advanced LIGO and LCGT are in the  
mirror curvature radius, beam radius, mirror material and temperature. 
\begin{table}[h]
\caption{\label{specification}Specification of Advanced LIGO 
\cite{Ju1,Ju2,Zhao} and LCGT \cite{Kuroda}.}
\begin{center}
\begin{tabular}{lll}
\br
 &Advanced LIGO&LCGT\\
\mr
Laser beam profile&Gaussian&Gaussian\\
Wavelength&1064 nm&1064 nm\\
Cavity length&4000 m&3000 m\\
Front mirror curvature radius&2076 m&7114 m\\
End mirror curvature radius&2076 m&7114 m\\
Beam radius at the mirrors&60 mm&35 mm\\
Power in a cavity &0.83 MW&0.41 MW\\
Mirror material&Fused silica&Sapphire\\
Mirror mass&40 kg&30 kg\\
Mirror temperature&300 K&20 K\\
\br
\end{tabular}
\end{center}
\end{table}

\subsection{Advanced LIGO}

We briefly overview an estimation of the instabilities in Advanced LIGO 
by a group 
at the University of Western Australia \cite{Ju1,Ju2} 
for an easier comparison. 
They investigated what happens when the curvature of 
a mirror is changed. The curvature of the other mirror is the default value 
given in Table \ref{specification}. Figure 1(c) of Ref. \cite{Ju2} shows the 
curvature dependence of the unstable mode number. 
The crosses in this figure represent 
the cavities, which consist of the fused-silica mirrors. 
The number of unstable modes is between 20 and 60.  
Figure 5 of Ref. \cite{Ju1} shows that the maximum of $R$ in the various 
elastic modes strongly depends on the mirror curvature.    
Even a shift of only a few meters in the mirror curvature 
causes a drastic change of the maximum $R$. The requirement of the accuracy 
in the mirror curvature in Advanced LIGO is difficult to be achieved.

\subsection{LCGT}

\begin{figure}[h]
\begin{minipage}{18pc}
\includegraphics[width=18pc]{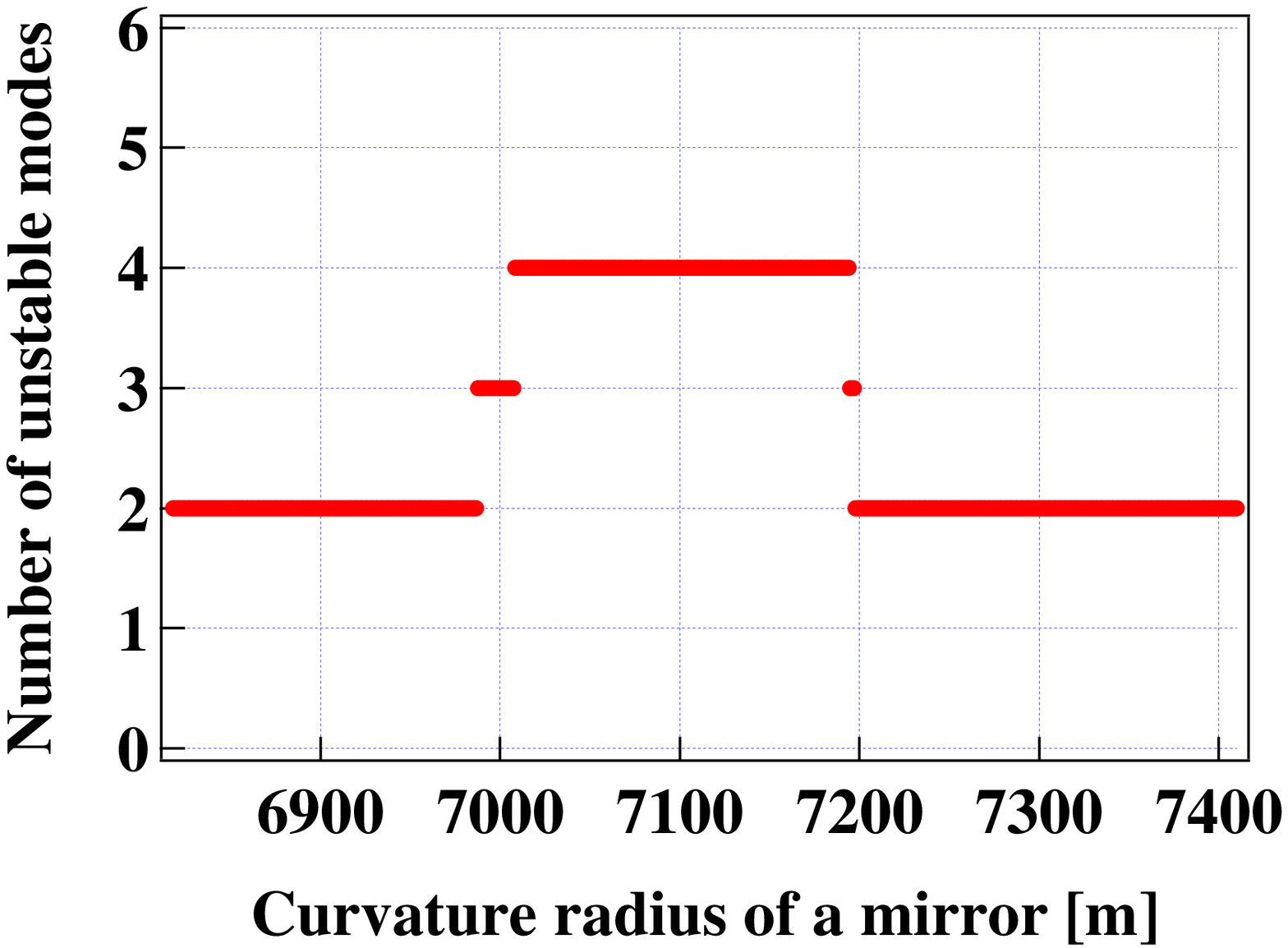}
\caption{\label{LCGT1}Number of unstable modes 
in the LCGT arm cavity. 
The horizontal axis is the curvature radius of a mirror. 
The curvature of the other mirror is the default 
value given in Table \ref{specification}. This graph corresponds to fig. 1(c) 
of Ref. \cite{Ju2} for Advanced LIGO.}
\end{minipage}\hspace{2pc}%
\begin{minipage}{18pc}
\includegraphics[width=18pc]{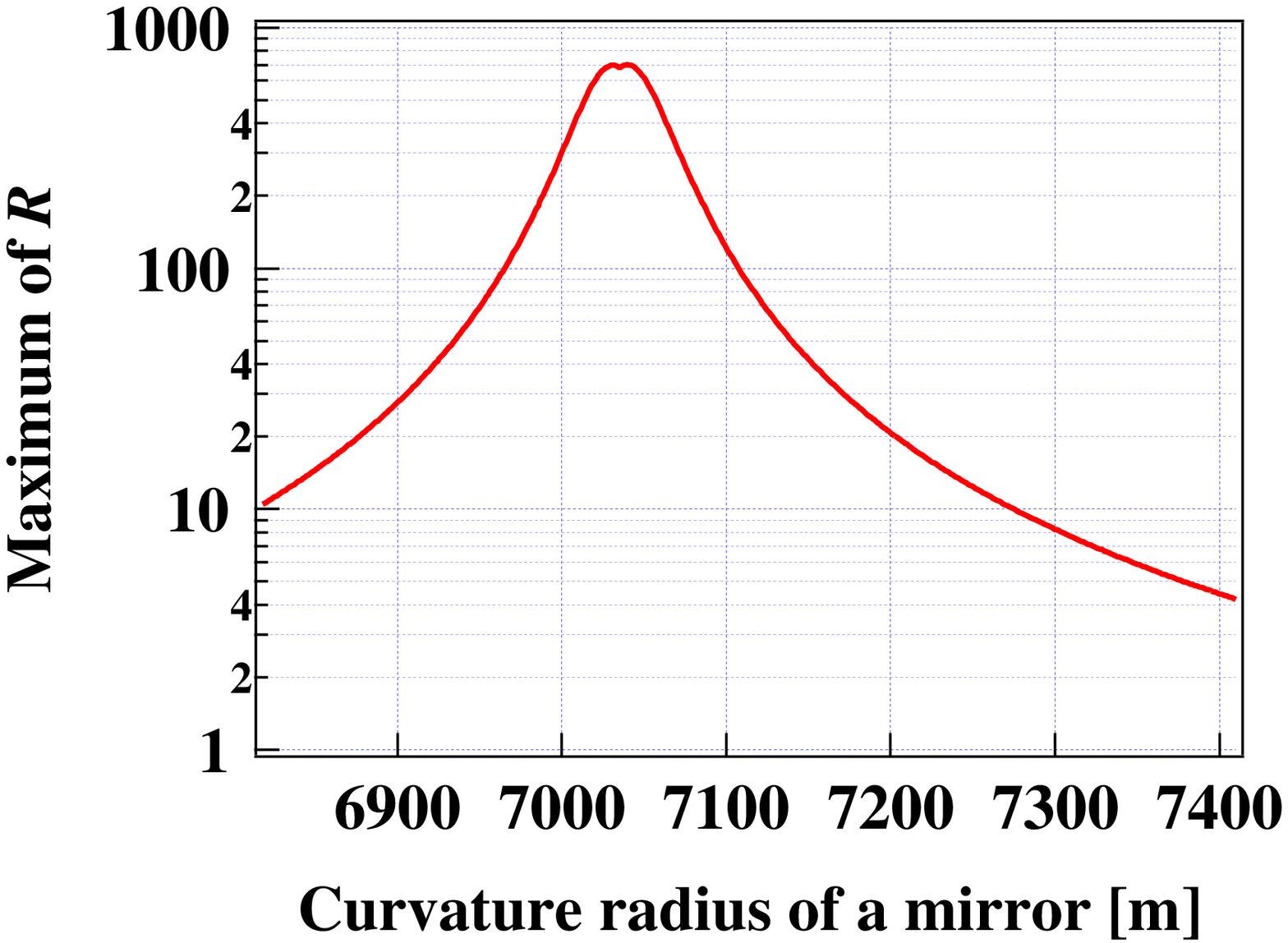}
\caption{\label{LCGT2}Maximum of $R$ in the various 
elastic modes in the LCGT arm cavity. The horizontal axis is 
the curvature radius of a mirror. 
The curvature of the other mirror is the default 
value given in Table \ref{specification}. 
This graph corresponds to fig. 5 
of Ref. \cite{Ju1} for Advanced LIGO.}
\end{minipage} 
\end{figure}

We investigated the parametric instabilities of the LCGT arm cavity 
in the same manner as 
that of the University of Western Australia. 
Figure \ref{LCGT1} shows the mirror curvature dependence of 
the unstable mode number in the LCGT arm cavity. 
The number is only 2 $\sim$ 4, which is 10-times smaller 
than that of Advanced LIGO.
Figure \ref{LCGT2} shows that the mirror curvature dependence of 
the maximum $R$ 
is weaker than that of Advanced LIGO.  
The maximum $R$ is not changed drastically 
by a shift of a few meters in the mirror curvature. 
It is easier to satisfy
the requirement of the mirror curvature in LCGT. 
It must be noticed that the power recycling was not taken into account 
in our calculation. The evaluation of Advanced LIGO 
included the power recycling effect. 
The difference between the maximums of $R$  
with and without the power recycling is shown in fig. 2 of Ref. \cite{Ju2}.
Although the peaks of the maximum $R$ 
become higher owing to the power recycling, 
this is not a serious effect on the discussion about the difference between 
Advanced LIGO and LCGT in the next section.  

\section{Discussion}

Our investigation revealed that there is the large difference in
the parametric instabilities
between Advanced LIGO and LCGT. We discuss the reasons in this section. 

\subsection{Number of unstable modes}

The difference in the unstable mode numbers originates 
from the mode frequency density difference. 
If both of the optical and elastic mode densities are large, the 
optical mode frequencies often coincide with the elastic mode frequencies.
The elastic mode density is inversely proportional to the cube of 
the sound velocity in the 
material. The sound velocities of the fused silica (Advanced LIGO) 
and sapphire (LCGT) 
are about 6 km/sec and 10 km/sec, respectively. 
The elastic mode density of LCGT is 5-times smaller. 
In Advanced LIGO, there are 7 optical transverse modes 
in a free spectrum range 
\cite{Ju1}. 
On the contrary, there are only 3 modes in the LCGT arm cavity. 
The optical mode density of LCGT 
is 2-times smaller. The larger optical mode density of 
Advanced LIGO stems from 
a larger beam radius adapted for suppressing the mirror thermal noise 
\cite{advancedLIGO}. 
In LCGT, since the mirrors are cooled in order to reduce 
the thermal noise \cite{LCGT}, larger beams  
are not necessary. As a result, the product of the elastic 
and optical mode densities of LCGT 
becomes 10-times smaller.
In fact, the unstable mode number calculated 
by us for LCGT (2 $\sim$ 4 in fig. \ref{LCGT1})
is 10-times less than that calculated 
at the University of Western Australia for Advanced LIGO  
(20 $\sim$ 60 in fig. 1(c) of Ref. \cite{Ju2}). 
 
\subsection{Mirror curvature dependence}

In LCGT, the maximum value of $R$ depends on the mirror 
curvature more weakly. 
This implies that the curvature dependence of the 
optical mode frequencies is weaker, 
because $R$ is a function of the optical mode frequencies. 
We calculated how the curvature variation affects the $n$-th optical 
transverse mode.
The results were 15$n$ Hz/m in Advanced LIGO and 0.58$n$ Hz/m in LCGT. 
LCGT shows a 30-times weaker dependence due to the larger optical 
transversal mode spacing, which stems from the smaller beam radius. 

\subsection{Summary of discussion}

The difference in the parametric instabilities between Advanced LIGO 
and LCGT is caused by those 
of the beam radii (Advanced LIGO, 60 mm; LCGT, 35 mm) 
and the mirror materials (Advanced LIGO, fused silica; LCGT, sapphire). 
These differences mostly originate from that of the 
thermal noise-reduction methods (Advanced LIGO, fused silica mirrors 
with larger laser beams; 
LCGT, cooling sapphire mirrors), 
which are the main strategies of the projects \cite{advancedLIGO,LCGT}. 
The cryogenic interferometer has an advantage in the 
parametric instabilities (less unstable mode number 
and weaker dependence on the mirror curvature) 
in addition to the small thermal noise 
\cite{Uchiyamasapphire,Uchiyamafiber,Yamamotocoating} 
and negligible thermal lensing \cite{TomaruAmaldi4}. 

\section{Instability suppression for LCGT arm cavity}

We showed that the unstable modes of LCGT 
are less than those of Advanced LIGO. 
Still, the LCGT arm cavity is unstable 
because there always exist unstable modes, 
as shown in fig. \ref{LCGT1}. We must consider 
how to suppress the instabilities. The three 
methods for instability suppression in Advanced LIGO are being 
studied at the University of 
Western Australia \cite{Ju2,GrasAmaldi6}. 
We checked whether these three methods (thermal tuning method, 
feedback control, Q reduction of elastic modes) are appropriate for LCGT 
(the tranquilizer cavity \cite{tranquilizer} 
is one of the other methods. However, this is difficult).

\subsection{Thermal tuning method}

In the thermal tuning method \cite{Ju2}, a part of the mirror is heated  
for curvature control. 
Since $R$ depends 
on the curvature, the suppression of $R$ should be possible by this manner. 
However, this method is not useful in LCGT.
First of all, $R$ weakly depends on the mirror curvature, as shown in 
fig. \ref{LCGT2} in the case of LCGT.  
Second, owing to the small thermal expansion 
and high thermal conductivity \cite{TomaruAmaldi4}
of cold sapphire, the mirror curvature 
would not change effectively, even if we apply heat to the mirror. 

\subsection{Feedback control}

It is possible to control the light 
or the mirror so that the parametric instabilities 
would be actively suppressed \cite{Ju2}. 
If the number of unstable modes is smaller, feedback control is 
easier. However, these are more difficult (active) methods 
than Q reduction (passive method) of the elastic modes. 

\subsection{Q reduction of elastic modes}

\begin{figure}[h]
\includegraphics[width=18pc]{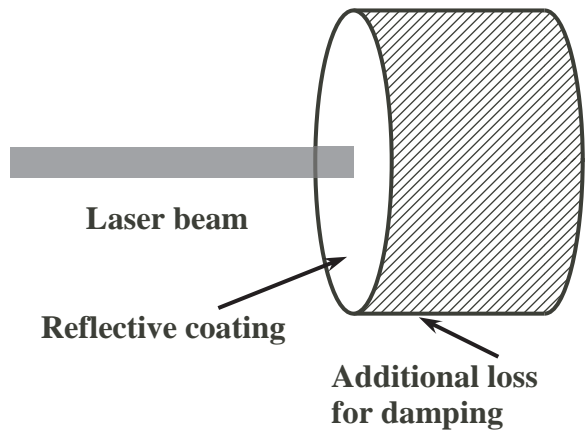}\hspace{2pc}%
\begin{minipage}[b]{18pc}\caption{\label{coating}
Loss on the barrel surface.  
Although this loss decreases the elastic Q-values of the mirror, 
$Q_{\rm m}$, it 
has only a small contribution to the thermal noise \cite{Levin,Yamamoto2}. 
Thus, this loss suppresses the parametric instabilities 
without an increase of the thermal noise \cite{GrasAmaldi6,Gras}.}
\end{minipage}
\end{figure}

This is a useful method \cite{GrasAmaldi6} for LCGT. 
The value of $R$ is proportional to the Q-value 
of the elastic mode, $Q_{\rm m}$, as shown in Eq. (\ref{R}). 
The Q-values of sapphire are about $10^8$ 
\cite{Uchiyamasapphire}. The maximum $R$ of LCGT is several hundreds
at most, as shown in fig. \ref{LCGT2}. 
If the Q-values of the LCGT mirrors become $10^6$, 
almost all modes become stable. 
Since the mechanical loss concentrated far from the beam spot 
has a small contribution to 
the thermal noise \cite{Levin,Yamamoto2}, 
we should be able to apply additional 
loss on a barrel surface, as in fig. \ref{coating},  
without sacrificing the thermal noise \cite{GrasAmaldi6,Gras}. 
Moreover, it is possible to introduce a 15-times larger loss 
in the LCGT mirror than that in Advanced LIGO mirror,  
because the LCGT mirrors are cooled (20 K) \cite{LCGT}. 
We concluded that 
the thermal noise of the barrel surface loss is comparable 
with that of the 
reflective-coating loss \cite{Yamamotocoating, Yamamoto2}, 
when the LCGT mirror Q-values become $10^6$ owing to the additional loss. 
Since the reflective-coating thermal noise is smaller 
than the goal sensitivity of LCGT \cite{Yamamotocoating}, 
this barrel surface loss is not a serious problem. 

We are able to introduce loss on the barrel surface by coating Ta$_2$O$_5$, 
which is a popular material 
for the reflective coating of the mirror. 
Our recent measurement \cite{Yamamotocoating} proved that the loss angle 
of the SiO$_2$/Ta$_2$O$_5$ coating is $(4 \sim 6) \times 10^{-4}$ 
between 4 K and 300 K. 
Since the loss of this coating is dominated by that of Ta$_2$O$_5$ 
\cite{Penn},    
the loss angle of Ta$_2$O$_5$ is $(8 \sim 12) \times 10^{-4}$. 
If the barrel loss dominates the mirror Q, 
it would be expressed as \cite{Yamamoto2}
\begin{equation}
\frac1{Q_{\rm m}} \sim \frac{E_{{\rm Ta}_2{\rm O}_5}}{E_{\rm sapphire}}
\frac{2 d}{R}\phi, 
\end{equation}
where $E_{{\rm Ta}_2{\rm O}_5},E_{\rm sapphire}, d, R, \phi$ are 
the Young's moduli of Ta$_2$O$_5$ and the sapphire, 
the thickness of the Ta$_2$O$_5$ layer, the mirror radius and the 
loss angle of Ta$_2$O$_5$, respectively. 
These values are summarized in Table \ref{coatingspe} 
\cite{Yamamotocoating}. 
In order to make the Q-values, $Q_{\rm m}$, $10^6$, 
the Ta$_2$O$_5$ coating thickness, $d$, 
must be 0.2 mm. 
\begin{table}[h]
\caption{\label{coatingspe}Specification of the coating 
\cite{Yamamotocoating}.}
\begin{center}
\begin{tabular}{ll}
\br
Young's modulus of the Ta$_2$O$_5$ ($E_{{\rm Ta}_2{\rm O}_5}$)
&$1.4 \times 10^{11}$ Pa\\
Young's modulus of the sapphire ($E_{\rm sapphire}$)&$4.0 \times 10^{11}$ Pa\\
Mirror radius ($R$)&12.5 cm\\
Loss angle of Ta$_2$O$_5$ ($\phi$)&$10^{-3}$\\
\br
\end{tabular}
\end{center}
\end{table}

\section{Future work}

In our calculations for this article, we took only the elastic modes below 
100 kHz and the first three transverse optical modes into account. 
We were calculating higher elastic and optical modes. 
Our preliminary result suggests that there are unstable 
higher modes. However, the elastic Q reduction would work more effectively 
because the typical $R$ still seems to be small. 
We must evaluate the Q reduction technique more carefully 
for instability suppression. 
The effects of power recycling \cite{PR}, 
resonant sideband extraction \cite{SR1,SR2} and the  
anti-Stokes modes \cite{anti-Stokes} must be evaluated as well.

\section{Summary}

We evaluated the parametric instabilities of LCGT 
and compared them 
with those of Advanced LIGO \cite{Ju1,Ju2}. 
The number of unstable elastic modes in LCGT 
is 10-times smaller.
Since the strength of the parametric instabilities 
in LCGT more weakly depends on the mirror curvature,  
the requirement of the accuracy in the mirror curvature of LCGT 
is easier to satisfy. 
These differences in the parametric instabilities 
between LCGT and Advanced LIGO are made by 
those of the laser beam sizes and mirror materials, 
which mostly stem from the thermal-noise suppression strategies 
in both projects 
(LCGT, cooling sapphire mirrors; 
Advanced LIGO, fused silica mirrors with larger laser beams) 
\cite{advancedLIGO,LCGT}. 
The elastic Q reduction \cite{GrasAmaldi6} 
by the barrel surface (0.2 mm thickness Ta$_2$O$_5$) coating 
is effective for instability suppression in the LCGT arm cavity. 
Thus, the cryogenic interferometer is a smart solution for the 
parametric instabilities in addition to the thermal noise 
\cite{Uchiyamasapphire,Uchiyamafiber,Yamamotocoating} 
and thermal lensing \cite{TomaruAmaldi4}. 

\ack

We are grateful to M Ando for information about candidates of 
the LCGT mirror curvature radius.


\end{document}